\documentclass[12pt]{article}
\usepackage{latexsym,amsfonts,amsmath,amsthm,amssymb}
\usepackage{pstricks}

\begin{document}

\title{Bell argument:  Locality or Realism?\\ Time to make the choice.}

\author{Andrei Khrennikov\\
International Center for Mathematical
Modelling \\ in Physics and Cognitive Sciences\\
Linnaeus University, S-35195, Sweden
\\Andrei.Khrennikov@lnu.se}

\maketitle

\begin{abstract}
This paper discusses a possible resolution of the nonobjectivity-nonlocality dilemma in quantum mechanics in 'the light of experimental tests of the Bell inequality for two entangled photons and a Bell-like inequality for a single neutron. My conclusion is that these experiments show that quantum mechanics is nonobjective: that is, the values of physical observables cannot be assigned to a system before measurement. Bell's assumption of nonlocality has to be rejected as having no direct experimental confirmation, at least thus far. I also consider the relationships between nonobjectivity and contextuality. Specifically, I analyze the impact of the Kochen-Specker theorem on the problem of contextuality of quantum observables. I argue that, just as von Neumann's ``no-go'' theorem, the Kochen-Specker theorem is based on assumptions that do not correspond to the real physical situation. Finally, I present a theory of measurement based on a classical, purely wave model (pre-quantum classical statistical field theory), a model that reproduces quantum probabilities. In this model continuous fields are transformed into discrete clicks of detectors. While this model is classical, it is nonobjective. In this case, nonobjectivity is the result of the dependence of experimental outcomes on the context of measurement, in accordance with Bohr's view.
\end{abstract}

\section{Introduction}

The common interpretation of J. Bell's argument is that a violation of his inequality implies that local realism has to be rejected. Experimental tests \cite{ASP} -- \cite{W2} can be considered as signs\footnote{Typically, the claim is stronger: it was ``experimentally proven''... However, there are some problems with this type of assertion. See, e.g., \cite{CONT} and references there.}
that local realism contradicts experimental data and therefore has to be rejected.  However, the notion of ``local realism'' is ambiguous. It should, I argue, be split into two unambiguous notions, ``realism'' and ``locality,'' which were indeed separated by Bell \cite{B}.
	My analysis of several well-known experiments compels me to conclude that quantum mechanics is nonobjective, in the sense that the values of physical observables cannot be assigned to a quantum system before measurement. Bell's assumption of nonlocality (in the sense of ``action at a distance,'' allowing for instantaneous physical connections between spatially separated events), as a possible alternative, has to be rejected in view of the fact that there is no experimental evidence to support it, while it is in conflict with relativity, which is well confirmed experimentally. 
	I also discuss the relationships between nonobjectivity and contextuality. In particular, I analyze the impact of the Kochen-Specker theorem on the problem of the contextual nature of quantum observables. My conclusion is that, similarly to the von Neumann ``no-go'' theorem, the Kochen-Specker theorem is based on several assumptions that do not correspond to the real physical situation considered in quantum mechanics.
	Finally, I present a theory of measurements for a classical wave or field model (pre-quantum classical statistical field theory) that reproduces quantum probabilities. In this model, continuous fields are transformed into discrete clicks of detectors. While this model is classical (as concerns the behavior of quantum objects), it is nonobjective. In this case nonobjectivity is the result of contextuality — the dependence of measurement outcomes on the context of measurement (in accordance with N. Bohr's views).

\section{A resolution of dilemma: nonlocality or nonobjectivity?}

 I begin with defining our main terms:

\medskip

(R) {\bf Realism}: The possibility of assigning to a quantum system the values of observable quantities before measurement\footnote{ In philosophical terms, this is not the only or the most rigorous definition of realism or, to begin with, reality or objectivity (objective reality). This is because it is more philosophically appropriate to assume that in order for something to be real or objective, or objectively real, it is sufficient for this something to exist independently, and hence, without any necessary relation with experiment. See the definition of value definiteness (VD) in the section on the Kochen-Specker theorem below. However, Bell used the term ``measurement realism'' in accordance with my definition here. Indeed, if the values of physical observables existed independently, but did not coincide with the outcomes of measurements, then Bell's argument would not imply Bell's inequality. See \cite{KV}--\cite{KV3} for an analysis and examples. In philosophic literature (R) is often referred as the principle of faithful measurement (FM) \cite{EE}. See, again, the discussion of the Kochen-Specker theorem in section 3.1.}  and these values are confirmed by this measurement. 

\medskip

(L) {\bf Locality}: No action at a distance. 

\medskip

Therefore, anyone who accepts that experiments are strong signs that local realism has to be rejected has to make the choice between:

(NONL) Realism, but nonlocality (Bell's position).
(NR) No realism (nonobjectivity) and locality (Bohr's position).
(NONL+NR) Nonlocality + nonobjectivity.

	The last possibility, (NONL+NR), appears to be too complex and is unlikely to occur in nature. One cannot of course completely exclude the possibility that nature is sufficiently exotic for this to happen. However, the (NONL+NR)-interpretation of experimental results can, at the very least, be excluded by the Occam-razor type of reasoning as a non-parsimonious option.

	In any event, one need not make this assumption in order to resolve the problem in question in this paper. It is sufficient to consider the alternative between nonlocality and nonobjectivity, and to choose between them. Thus, the choice is between the position taken by Bell (or by D. Bohm, at least, insofar as Bohmian mechanics it concerned) or the position taken by Bohr, and, following him, W.Heisenberg and W. Pauli. It should be noted that the Copenhagen interpretation of quantum mechanics, at least as understood by the latter three figures, rejects nonlocality. Bohr maintained that the values of observable quantum quantities are ``created'' in the process of the interaction between quantum systems and measurement devices. Accordingly, his main point was the impossibility of realism in the sense of assigning independently existing, ``objective,'' properties to quantum objects themselves. 

	It is typically assumed that the present experimental situation in quantum physics does not provide us with a possibility to make the choice between the two alternative positions in question; this is indeed correct if one only considers experiments of the EPR-Bohm type, in which realism and locality are intermixed. However, an exciting recent experiment in neutron interferometry supports the thesis that quantum mechanics is contextual \cite{R}. But contextuality implies nonobjectivity!\footnote{I reiterate that by objectivity or realism I understand ``measurement'' objectivity or realism. See note 2 above. Accordingly, contextuality does not imply the violation of the principle of value definiteness (VD).} In the contextual situation, it is impossible to assign the values of physical observables before measurement. Therefore, this experiment can be considered as supporting nonobjectivity. It is an experiment about the nonobjectivity of the results of measurements for a single particle.

	The following assumption, then, would seem to be logically justified. If even the case of a single particle already exhibits a lack of objectivity in quantum physics, then it is reasonable to assume that the situation cannot be improved by considering a pair of particles. Hence, it is also reasonable to expect and even assume nonobjectivity in the EPR-Bohm experiment. This implies that in the alternative (NONL) or (NR), the choice of (NR) is more justified than the choice of (NONL).  A following conclusion may then be drawn:  {\it Recent experiments in foundational aspects of quantum physics can be considered as supporting Bohr's view that quantum observables are nonobjective: their values cannot be assigned before measurement. The assumption of nonlocality has to be rejected, because there is no direct experimental evidence of nonlocality, comparable to the test of nonobjectivity performed in \cite{R}, and because in the EPR-Bohm type of experiments it is not necessary to assume nonlocality, if one assumes nonobjectivity. }

	Two additional remarks are in order at this point.  First of all, because, the experiments with neutrons have practically 100\% efficiency, the experiment of H. Rauch  and his team just discussed can be considered as essentially loophole free.	Secondly, Bohr's interpretation of quantum mechanics and his position on the situation in question was elaborated through an analysis of the physical and philosophical consequences of Heisenberg's uncertainty principle. This justification of nonobjectivity of quantum mechanics was later strongly criticized by H. Margenau \cite{M} and L. Ballentine \cite{BL}, considered in detail in an earlier book by the present author \cite{KV3}. They rightly pointed out that Heisenberg's uncertainty principle could only be rigorously presented in the form of the Schr\"odinger-Robertson inequality for standard deviations for two incompatible (operator) observables. However, standard deviations are statistical quantities and each of them is calculated independently from another, i.e., we can first perform a series of the position measurements and find sx and then perform independently the momentum measurements and find sp. Accordingly, it appears impossible to justify Heisenberg's original position that his inequality implies nonobjectivity. Therefore, the test of contextuality considered above \cite{R} indeed plays a crucial role in the justification of Bohr's position.

\subsection{Against nonlocality}

The present argument actually follows a widespread earlier view of Bell's theorem and even of the EPR experiment. For example, in his biography of Einstein, A. Pais \cite{Pais} says flatly that there is no paradox in the EPR experiment, and that the latter only means that quantum mechanics is not objective. A. Plotnitsky, in a long note addressing the subject in his recent book \cite{P}, note 8, p. 247, says that nonlocality of QM is a minority view. Kurt Gottfried's article, mentioned by Plotnitsky, expressly states that in relativity we in fact have a test that rules out nonlocality. C. A. Fuchs, on the other hand, thinks \cite{Fuchs2} that by now nonlocality of quantum mechanics is the majority view, while the kind of view advocated in this paper or his view (which is somewhat different, since he takes a subjectivist view of both quantum states and of probability itself \cite{Fuchs}--\cite{Fuchs1}) is a minority view now.
	
I am not certain that this is necessarily true, once one considers the physics community as a whole (and not only its quantum information part). However, it is, in my view, not easy, if possible at all, to change the view of those who believe that nonlocality is a consequence of Bell's theorem or of the Kochen-Specker theorem, in part because there are also arguments that the nonlocality in question cannot be detected experimentally and, hence, there is no violation of relativity in practical terms.

\section{Contextuality}

\subsection{Kochen-Specker theorem}

In this paper, I am concerned not with the mathematics or physics related to the Kochen-Specker theorem, but with philosophic considerations surrounding it (e.g., \cite{EE}.) The explicit premise of the hidden variables (HV) interpretation of theories, as understood in literature on quantum foundations, is one of value definiteness (VD):

\medskip

(VD) {\it All observables defined for a QM system have definite values at all times.}

\medskip

According to H. Carsten:

(VD) is motivated by a more basic principle, an apparently innocuous realism about physical measurement which, initially, seems an indispensable tenet of natural science. This realism consists in the assumption that whatever exists in the physical world is causally independent of our measurements which serve to give us information about it. Now, since measurements of all QM observables, typically, yield more or less precise values, there is good reason to think that such values exist independently of any measurements - which leads us to assume (VD). Note that we do not need to assume here that the values are faithfully revealed by measurement, but only that they exist! \cite{EE}. 

It follows that our assumption of realism (or objectivity), (R), is stronger than (VD), i.e., (R) implies (VD). By (R), the values of physical observables not only exist, but they are also faithfully revealed by measurement. (R) is also known as the principle of faithful measurement (FM) \cite{EE}.
	
One can concretize our ``innocuous'' realism by the second assumption of non- contextuality:

\medskip

	(NC) {\it If a QM system possesses a property (value of an observable), then it does so independently of any measurement context, i.e. independently of how that value is eventually measured.}
	
\medskip	

	By the Kochen-Specker theorem (ND) and (NC), under the additional assumptions  of the sum rule and the product rule \cite{EE}, are incompatible with quantum mechanics. I restate these rules here. Values of observables conform to the following constraints:

\medskip	
	
(a) If $A, B, C$ are all compatible and $C = A+B,$ then $v(C) = v(A)+v(B);$ 

\medskip	

(b) if $A, B, C$ are all compatible and $C = A•B,$ then $v(C) = v(A)•v(B).$

\subsection{Contextuality}

Since the notion of contextuality used in literature on quantum foundations often depends on the problems one considers and, hence, vary, I would like to follow the philosophical definition that expresses the general content of this notion and not its specific applications, (cf. \cite{B}, \cite{KV3}). The negation of (NC) gives us the following general definition of contextuality:

\medskip	

(C) {\it If a QM system possesses a property (a value of an observable), then it does so depending on the concrete measurement context, i.e. depending on how that value is eventually measured.}

\medskip

	This definition of contextuality is close to the view of Bohr \cite{BR},  \cite{BR0} who often emphasized that the whole context of measurement has to be taken into account, although, as I explain below, for Bohr the ``system'' would be constituted by a certain indivisible wholeness of the quantum object considered and the measuring apparatus involved (cf., also \cite{KV3}). On the other hand, this definition is more general than Bell's definition \cite{B} of contextuality, used in works devoted to the contextual analyses of Bell's inequality and other ``no-go'' statements. In this view, that context of measurement of an observable $A$ is reduced to the presence of other observables compatible with $A.$ (Bell defined noncontextuality in the following way: ``measurement of an observable must yield the same value independently of what other measurements may be made simultaneously'' \cite{B}, p. 9.) My opinion (which, again, coincides with, or at least is close to that of Bohr) is that Bell's contextuality has no direct relation to the real contextuality of quantum observables – the contextuality of measurement of a single observable, having no direct relation to the presence or absence of other observables compatible with A. On the other hand, it is clear that, as things stand now, only Bell's contextuality has relation to real experiments. All known tests of contextuality concern Bell's contextuality. It is not clear how to test experimentally the fundamental contextuality of a single value of a single observable. Developing such a test would essentially clarify the problem of contextuality in quantum mechanics.
	
I would like to add that Bell's view of contextuality was actually more general than the assumption of the contextuality of joint measurements. In particular, he wrote:

\medskip

``A final moral concerns terminology. Why did such serious people take so seriously axioms which now seem so arbitrary? I suspect that they were misled by the pernicious misuse of the word 'measurement' in contemporary theory. This word very strongly suggests the ascertaining of some preexisting property of some thing, any instrument involved playing a purely passive role. Quantum experiments are just not like that, as we learned especially from Bohr. The results have to be regarded as the joint product of system and apparatus, the complete experimental set-up. But the misuse of the word 'measurement' makes it easy to forget this and then to expect that the results of measurements should obey some simple logic in which the apparatus is not mentioned. The resulting difficulties soon show that any such logic is not ordinary logic. It is my impression that the whole vast subject of Quantum Logic has arisen in this way from the misuse of a word. I am convinced that the word 'measurement' has now been so abused that the field would be significantly advanced by banning its use altogether, in favor for example of the word ``experiment''.'' \cite{B}

This supports Bohr's viewpoint of the role of observables in QM. In addition to the experiments in neutron interferometry mentioned above, which test contextuality with the aid of Bell's inequality, one can also mention the experiments testing the assumption of contextuality in the framework of the Kochen-Specker theorem and its generalizations \cite{AD}. However, contextuality in the Kochen-Specker arguments is mixed with other assumptions, some of which are clearly nonphysical (i.e., they do not correspond to any actual physical situation), as are, for example, the sum rule and the product rule \cite{EE}. Therefore such tests cannot be considered as tests of ``pure noncontextuality''.

	Rules (a) and (b) in the Kochen-Specker theorem are natural, for example, for classical phase-space mechanics where observables are given by functions on the phase-space. However, it is not clear why they should hold for any prequantum model. In general, the ``no-go'' arguments seem to be directed not against all possible ``prequantum models'' reproducing quantum predictions, but only against classical statistical mechanics. Considered from this restricted viewpoint, the conclusions of these arguments could be seen as valid. The problem is that those who advance such arguments claim more than the known no-go theorems in fact imply. I cannot describe this situation better than Bell himself did: ``long may Lois De Broglie continue to inspire those who suspect that what is proved by impossibility proofs is 'lack of imagination'' \cite{B}.

\section{Contextuality and objectivity}

	The postulates (VD) (ontic realism) and (NC) are logically independent. Therefore (VD) can survive even in the (C)-world. On the other hand, (R) (also known as (FM)) and (NC) are logically dependent. Therefore, the survival of (R) in the (C)-world is questionable. This is a complicated question. In principle, a possibility that (R) and (C) could coexist is not logically excluded. The detailed analysis of this problem is presented in \cite{EE}, where it is rightly pointed out that, although the coexistence of (R) and (C) cannot be completely excluded, any attempt to imagine more or less natural realization of contextuality in an experiment leads to a rejection of (R). Following \cite{EE}, we consider several types of (C) related to measurement.

\subsection{Causal contextuality}

	An observable might be causally context-dependent in the sense that it is causally sensitive to how it is measured. The basic idea here is that the observed value comes about as the effect of the system-apparatus interaction. Hence, measuring a system via interaction with an apparatus measuring $P$ might yield a value $v(P),$ while measuring the same system via interaction with an apparatus measuring $Q$ might yield a different value $v(Q),$ although both observables are represented by the same operator (quantum observable). The difference in values is explained in terms of the context-dependence of the observables: The latter are context-dependent because the different ways of physically realizing them causally influence the system in different ways and thereby change the observed values. The usage of causal contextuality is definitely incompatible with (R), but (VD) can still be considered as a reasonable assumption. Thus, ``ontic realism'' survives, but ``measurement realism'' does not.

\subsection{Ontological contextuality}

	An observable might be ontologically context-dependent in the sense that, in order for it to be well defined, the specification of the observable that it comes from is necessary. Any attempt to create an experimental picture of ontological contextuality would generate a diversity of opinions and pictures \cite{EE}. We are interested in the following picture:

	Any property, rather than being dependent on the presence of another property, is dependent on the presence of a measuring apparatus. This amounts to a holistic position: Some properties could only be meaningfully considered as pertaining to the system, if that system is part of a certain system-apparatus whole. This viewpoint is strongly reminiscent of that of Bohr, as developed in his 1935 response to EPR \cite{BR}.

\section{Death of hidden variables and born of subquantum variables}

The present author spent 18 years working on quantum foundations, and the final conclusion that emerged from this long effort is that HV should be rejected. It was not an easy decision, as can be seen from my earlier (2001) V\"axj\"o interpretation of QM \cite{VI1}, which was an (NC)+(R) interpretation. My decision to abandon HV was not a consequence of my better understanding of no-go theorems. (The better I understand them the more problems I see in their assumptions, especially in matching these assumptions to the real experimental situation \cite{KV3}.) I am still convinced that the Bell theorem collapses when confronting the problem of efficiency of detectors or, more generally, unfair sampling \cite{US}, \cite{US1} (including the experimentally important version of unfair sampling based on the usage of the time window \cite{Raedt}). I am still convinced that the von Neumann and Kochen-Specker theorems do not have much to do with the real experimental situation, and remain merely mathematical exercises \cite{KV3}. However, through the study of Bohr's works and his interpretation of quantum observables as representing measurements related to various contexts, I came to the conclusion that ``naive Einsteinian realism,''(R) has to be rejected, that (NC) has to be rejected, and that QM is contextual \cite{VI2}. However, I thought that (VD) could still survive.

	Recently, I developed a new purely wave model (prequantum classical statistical field theory, PCSFT) \cite{PC1}-\cite{PC3} which reproduces the main probabilistic predictions of QM, including correlations of entangled systems. However, the correspondence between observables in PCSFT and QM was rather tricky. PCSFT is not a theory of HV for QM in the traditional sense. PCSFT has its own basic variables and fields' coordinates, $\phi=(\phi_j).$ However, because the values of standard quantum observables cannot be assigned to such ``sub-quantum variables,'' both postulates of the conventional HV-theory, (VD) or value definitiveness and (R)/(FM) or measurement realism, are violated. The measurement theory for PCSFT is contextual, and this contextuality is of the type considered by Bohr, in accordance with the above discussion of ontological contextuality: it only makes sense to speak of quantum observables as pertaining to the system if that system is part of a certain system-apparatus whole. The subquantum field-type variable $\phi$ plays a crucial role in the creation of values of quantum observables, as clicks of detectors. However, the functional representation of quantum observables, $\phi \to A(\phi)$ is impossible.
	
The temporal structure of the measurement process plays a fundamental role (cf. \cite{HP}-- \cite{HP2}). In fact, the subquantum variables determine only the instant of a detector's click. Thus, for a fixed instant of time, it is impossible to determine the values of all possible quantum observables, even of any two of them, and even when they are compatible observables. The measurement theory of PCSFT matches so well with the Bohr's view that one might even imagine that Bohr could have rejected his postulate on completeness of QM in favor of such a contextual model with subquatum variables.
	
I now move to a brief exposition of measurement theory of PCSFT \cite{PC1}--\cite{PC3}, which I develop in detail in \cite{MM}. To reiterate, PCSFT is part of the classical theory of signals. It treats a special class of random signals (with covariance operators of a special type) and a special class of observables for classical signals (given by quadratic forms). The tricky point is the correspondence between PCSFT-variables, classical field variables $\phi,$ and quantum observables. The latter are represented by clicks of detectors. It is crucial that our description of the measurement process is based on the presence of two time scales: a) the prequantum time scale – the scale of fluctuations of the classical field which is symbolically represented as a quantum particle, and this scale is very fine; b) the scale of quantum measurements, and this scale is very coarse in comparison with the prequantum scale. Relative to the prequantum time scale, quantum measurement takes a very long time, in this mathematical model – practically, infinitely long. The values of quantum observables are created through such a process, in accordance with the concept of causal contextuality. By moving from the prequantum time scale to the scale of quantum measurements we determine instances of clicks, the frequency of clicks for the values of conventional quantum observables, and the probabilities of these values.

\section{Random signals}  

The state space of classical signal theory  is the $L_2$-space $H=L_2({\bf R}^3).$ Elements of 
$H$ are classical fields $\phi: {\bf R}^3 \to {\bf C}^n.$ We consider complex valued fields; for
example, for the classical electromagnetic field we use Riemann-Silberstein representation,
$\phi(x) = E(x) +i B(x).$\footnote{Later we shall move from the general theory of classical random signals to PCSFT and then to QM. A consideration of the complex representation of classical fields induces the usage of complex numbers in QM. Thus, in the present approach there is nothing mystical in the presence of a complex-number structure of QM, in particular, in the fact that one can derive probabilities from complex amplitudes.} A random field (signal) is a field (signal) depending on a random parameter $\omega, 
\phi(x, \omega).$ In the measure-theoretic framework (Kolmogorov, 1933) it is represented as 
$H$-valued random variable, $\omega \to \phi(\omega)\in H.$ Its probability distribution is denoted by the symbol 
$\mu$ on $H.$  Consider the functionals of fields, $f: H \to {\bf C}, \phi \to f(\phi).$ These are physical observables for classical signals.
For example, the energy of the classical electromagnetic field is geven by the quadratic functional
$$
f(\phi)\equiv f(E, B)= \int_{{\bf R}^3} \vert \phi(x) \vert^2 dx=  \int_{{\bf R}^3} (E^2(x) + B^2(x)) dx.
$$
The average of an  observable can be written as the integral over the space of fields
$$
\langle f\rangle = \int_H f(\phi) d \mu(\phi).
$$
To find $\langle f\rangle,$ we consider an ensemble (in theory, infinite) of realizations of the 
random field and calculate the average of $f(\phi)$ with respect to this ensemble. This measure-theoretic 
(ensemble) representation is very convenient in theoretical considerations \cite{S1}, \cite{S2}. However, in practice
we never produce an ensemble of different realizations of a signal.  Instead, we have a single time
dependent realization of a signal, $\phi(s,x).$ It is measured at different instances of time. Finally, we calculate 
the time average. The latter is given by 
\begin{equation}
\label{eqg1}
\bar{f}= \lim_{\Delta \to \infty} \frac{1}{\Delta} \int_0^\Delta f(\phi(s)) ds.
\end{equation}
In classical signal theory \cite{S1}, \cite{S2} the ensemble and time averages are coupled by
the {\it ergodicity assumption.} Under this assumption we obtain that 
\begin{equation}
\label{eqg2}
\bar{f}= \langle f \rangle,
\end{equation}
i.e., 
\begin{equation}
\label{eqg3}
\int_H f(\phi) d \mu(\phi) = \lim_{\Delta \to \infty} \frac{1}{\Delta}\int_0^\Delta f(\phi(s)) ds \approx 
\frac{1}{\Delta}\int_0^\Delta f(\phi(s)) ds,
\end{equation}
foe sufficiently large $\Delta.$ 

From this point on, we shall operate only with observables given by quadratic functionals of 
classical signals:
\begin{equation}
\label{eqg4}
\phi \to f_A(\phi) = \langle \widehat{A} \phi, \phi \rangle,
\end{equation}
where $\widehat{A}$ is a self-adjoint operator.  Moreover, to describe a procedure of the position detection we  need only functionals of the form
\begin{equation}
\label{eqg5}
\phi \to \vert \phi(x_0)\vert^2, 
\end{equation}
where $x_0 \in {\bf R}^3$ is a fixed point that determines the quadratic functional
(later $x_0$ will be considered as the position of a detector).\footnote{
For the moment, my analysis remains within the general framework of theory of random signals. Later, as I move to my prequantum model, PCSFT, I shall consider random signals as representing quantum systems. The PCSFT quantities, (\ref{eqg4}), (\ref{eqg5}), do not directly belong to the domain of QM.}

In what follows we consider only random signals with covariance operators of the type 
\begin{equation}
\label{eqg41}
D_\psi= \vert \psi\rangle \langle \psi \vert,
\end{equation}
 where $\psi \in H$ is arbitrary vector 
(i.e., it need not be normalized by 1).\footnote{This is just a special class of classical random signals. In PCSFT
such signals will represent quantum systems in pure states.}
For such $\mu\equiv \mu_\psi,$ 
\begin{equation}
\label{EQj1}
\langle f_{x_0}\rangle = \int_H \vert \phi(x_0)\vert^2 d\mu_\psi(\phi) = \vert \psi(x_0)\vert^2.
\end{equation}
And under the assumption of ergodicity, we obtain
\begin{equation}
\label{EQj2}
  \vert \psi(x_0)\vert^2= \lim_{\Delta \to \infty} \frac{1}{\Delta} \int_0^\Delta \vert \phi(s, x_0)\vert^2 ds 
\approx \frac{1}{\Delta} \int_0^\Delta \vert \phi(s, x_0)\vert^2 ds,
\end{equation}
for sufficiently large $\Delta.$
Consider the functional 
\begin{equation}
\label{EQj3}
\pi(\phi) = \Vert \phi \Vert^2= \int_{{\bf R}^3} \vert \phi(x)\vert^2  dx.
\end{equation}
In PCSFT, it represents the total energy of a signal.(However, this is not the conventional quantum observable.
It is an internal quantity of PCSFT. To obtain conventional quantum quantities, we have to perform detections,
which will be considered in the next section.) We find its average. In general,
\begin{equation}
\label{EQj4}
\langle \pi\rangle = \int_H \pi(\phi) d\mu(\phi) = \rm{Tr} D_\mu.
\end{equation}
In particular, for $\mu=\mu_\psi,$ 
\begin{equation}
\label{EQj5}
\langle \pi\rangle = \int_H \pi(\phi) d\mu_\psi(\phi) = \Vert \psi \Vert^2.
\end{equation}
By ergodicity
\begin{equation}
\label{EQj6}
\langle \pi\rangle = \Vert \psi \Vert^2 = \lim_{\Delta \to \infty} \frac{1}{\Delta} \int_0^\Delta \Vert \phi(s)\Vert^2 ds
\approx \frac{1}{\Delta} \int_0^\Delta ds \int_{{\bf R}^3}  dx \vert \phi(s,x)\vert^2,
\end{equation}
for sufficiently large $\Delta.$

If, as usual in signal theory, the quantity $\vert \phi(s,x)\vert^2$ has the physical dimension 
of the energy density, i.e., energy/volume, then by selecting some unit of time denoted $\gamma$
we can interpret the quantity 
 \begin{equation}
\label{EQj7}
\frac{1}{\gamma} \int_0^\Delta \vert \phi(s, x_0)\vert^2 ds dV,
\end{equation}
as the energy which can be collected in the volume $dV$ during the time interval $\Delta$ (from the random 
signal $\phi(s) \in H).$ In the same way 
\begin{equation}
\label{EQj8}
\frac{1}{\gamma} \int_0^\Delta ds \int_{{\bf R}^3} dx \vert \phi(s, x)\vert^2,
\end{equation}
is the total energy which can be collected during the time interval $\Delta.$ 
Its time average can be represented in the form (\ref{EQj6}).

\section{Discrete-counts model for detection of classical random signals}
\label{mesj}

We consider the following model of a detector's functioning. Its basic parameter is detection 
threshold energy $\epsilon\equiv \epsilon_{\rm{click}}.$ The detector under consideration clicks
after it has collected the energy 
\begin{equation}
\label{EQj9}
E_{\rm{collected}} \approx \epsilon.
\end{equation}
Such a detector is calibrated to work in accordance with (\ref{EQj9}). Realizations of the random signal with energies deviating from 
$\epsilon$ are discarded. Detectors are calibrated for a class of signals and the corresponding $\epsilon$ is selected. 
Let us select $\gamma,$ as  one second. Consider such a detector located in a small volume $dV$ around a point  $x_0 \in {\bf R}^3.$ 
In average it clicks each  $\Delta$ seconds, where $\Delta$ is determined from the approximative equality
\begin{equation}
\label{EQj11}
\frac{1}{\gamma} \int_0^\Delta \vert \phi(s, x_0)\vert^2 ds dV\approx \epsilon,
\end{equation}
or 
\begin{equation}
\label{EQj12}
\frac{\Delta}{\gamma} \Big( \frac{1}{\Delta} \int_0^\Delta \vert \phi(s, x_0)\vert^2 ds\Big) dV\approx \epsilon,
\end{equation}
or 
\begin{equation}
\label{EQj13}
\frac{\Delta}{\gamma} \vert \psi(x_0)\vert^2 dV \approx \epsilon.
\end{equation}

Thus at the point $x_0$ such a detector clicks (in average) with the frequency
\begin{equation}
\label{EQj15}
\lambda(x_0) = \frac{\gamma}{\Delta} \approx \frac{\vert \psi(x_0)\vert^2 dV}{\epsilon}. 
\end{equation}
This frequency of clicks coincides with the probability of detection at the point $x_0.$  
Consider a large interval of time, say $T.$ The number of clicks at $x_0$ during this interval 
is given by
\begin{equation}
\label{EQj16}
n_T(x_0) =\frac{T\vert \psi(x_0)\vert^2 dV}{\epsilon \gamma}. 
\end{equation}
The same formula is valid for any point $x \in {\bf R}^3.$ Hence, the probability of detection at $x_0$
is 
\begin{equation}
\label{EQj17}
P(x_0) = \frac{n_T(x_0)}{\int n_T(x) dx} \approx \frac{\vert \psi(x_0)\vert ^2 dV}{\int \vert \psi(x)\vert^2 dx}=
\vert \Psi(x_0)\vert ^2 dV,
\end{equation}
where the normalized function  
\begin{equation}
\label{EQj17pw}
\Psi(x)= \psi(x) /\Vert \psi \Vert,
\end{equation}
 i.e., $\Vert \Psi \Vert^2=1.$

Here $\Psi(x)$ is a kind of the wave function, a normalized vector of the $L_2$-space. (Once again, we still only consider classical signal theory.)

{\bf Conclusion.} {\it Born's rule is valid for probabilities of ``discretized detection'' of classical
random signals under the following assumptions:

(a) ergodicity;

(b) a detector clicks after it ``has eaten'' approximately a portion of energy $\epsilon;$

(c) the energy is collected by this detector through time integration of signal's energy;

(d) the interval of integration $\Delta$  is long enough from the viewpoint of the 
internal time scale of a signal.}

\medskip

The assumption (d) is necessary to match (a). I note that the 
internal time scale of a signal, i.e., the scale of its random fluctuations, 
 has to be distinguished from the time scale of macroscopic measurement (observer's 
 time scale). The former is essentially finer than the latter.
 
 The scheme just outlined is a natural scheme of discrete detections which is based on time 
integration of signal's energy by a detector. The calibration of the detector  plays a
crucial role. This scheme applied to classical random signals reproduces Born's rule 
for {\it discrete clicks.}

How can this detection scheme be applied to QM?

\section{Quantum probabilities from measurements of prequantum random fields}
\label{QP}

In PCSFT, quantum systems are represented by classical random fields. Hence, quantum measurements 
have to be interpreted as measurements of classical random signals. I shall now explore the measurement scheme 
of the previous section. 
Take a prequantum random field (signal) $\phi$ with zero average and the covariance operator given by 
(\ref{eqg41}): $D_\psi= \vert \psi\rangle \langle \psi \vert.$ Then we can introduce the wave function $\Psi$
by normalization of $\psi,$ see (\ref{EQj17pw}). We now consider quantum measurements for systems in the pure state
$\Psi$ as measurements of the corresponding classical signal $\phi$ and we derive the Born's rule for QM.

Thus, we arrived at a model of discrete detection of 
 prequantum  random fields (corresponding to quantum systems)  
 which reproduces the basic rule fo QM, Born's rule.

We stress that the resulting probability, see (\ref{EQj17pw}),   derived from PCSFT does not depend on the threshold $\epsilon,$  which is natural, since the  formula thus derived is nothing other than Born's rule. However, the frequency of clicks per time unit, $\lambda(x_0),$ depends inversively on $\epsilon,$ see (\ref{EQj15}).

\section{No double clicks}
\label{CAL1}

We recall that Bohr elaborated his complementarity principle\footnote{This principle is often called
``wave-particle'' duality. However,  Bohr  never used the latter terminology himself.}
from analysis of the two slit-experiment. On the one hand, quantum systems exhibit interference properties
which are similar to properties of classical waves. On the other hand, these systems also exhibit particle 
properties. Wave properties (interference) are exhibited if both slits are open and experimenter does not 
try to control 
through which slit particles pass, and when a sufficiently large number of particles hit the screen, where the interference pattern is registered. In this experimental context, one can be totally fine with a classical wave type model. However, if the experimental context is changed and detectors are placed behind the slits, then ``wave features of quantum systems disappear and particle features are exhibited.'' What does the latter fact mean? Why is the usage of the wave picture impossible? Typically, it is claimed that, since a classical wave is spatially extended, two detectors (behind both slits) can click simultaneously and produce double clicks. However, as it is commonly claimed, there are no double clicks at all; hence, the wave model has to be rejected (in the context of the presence of detectors). Bohr had not found any reasonable explanation of the context dependent features of quantum systems, and he elaborated the complementarity principle in order to consistently interpret this situation.
	Of course, the claim that there are no double clicks at all is meaningless at the experimental level. There are always double clicks. The question is whether the number of double clicks is very small (compared with the numbers of single clicks). The corresponding experiments have been performed [42], [43], which show that the number of double clicks is relatively small. Such experiments are considered as a confirmation of Bohr's complementarity principle.

I would argue that the absence of double clicks might not be fundamental, but is instead a consequence of the procedure of calibration of detectors. Consider again a random signal $\phi.$ But now we take two threshold
type detectors located in neighborhoods $V_{x_0}$ and $V_{y_0}$ of the points $x_0$ and $y_0.$ Suppose that both detectors have the same detection threshold $\epsilon.$ It is convenient to represent $\epsilon$ in the form $\epsilon = C \Vert \psi(x)\Vert^2,$ where the vector 
$\psi$ determines the covariance operator of the prequantum random signal and $C>0$ is a constant. (Here $\Psi=\psi/\Vert \psi\Vert$ is the quantum state corresponding the prequantum signal.) 
 For the moments of clicks, we have two approximate equalities:
 \begin{equation}
\label{EQj11D}
\frac{1}{\gamma} \int_0^{\Delta_C(x_0)} \int_{V_{x_0}} \vert \phi(s, x)\vert^2dx ds \approx C \Vert \psi(x)\Vert^2,
\end{equation}
 \begin{equation}
\label{EQj11D1}
\frac{1}{\gamma} \int_0^{\Delta_C(y_0)} \int_{V_{y_0}} \vert \phi(s, x)\vert^2  dx ds\approx C \Vert \psi(x)\Vert^2,
\end{equation}
A double click corresponds to the (approximate) coincidence of moments of clicks
 \begin{equation}
\label{EQj11D2}
\Delta_C(x_0, y_0) = \Delta_C(x_0) =\Delta_C(y_0).
\end{equation}
Hence, by adding the approximate equalities (\ref{EQj11D}), (\ref{EQj11D1}) under condition (\ref{EQj11D2}) we obtain
 \begin{equation}
\label{EQj11DD}
\frac{1}{\gamma} \int_0^{\Delta_C(x_0, y_0)} \int_{V_{x_0}\cup V_{y_0}} \vert \phi(s, x)\vert^2dx ds \approx 2 C\Vert \psi(x)\Vert^2,
\end{equation}
Again by using ergodicity and the assumption that the internal time scale of signals is essentially finer than
the time scale of measurement (``click production'') we obtain
$$
\frac{\Delta_C(x_0,y_0)}{\gamma}\Big[ \frac{1}{\Delta_C(x_0,y_0)} \int_0^{\Delta_C(x_0,y_0)} \int_{V_{x_0}\cup V_{y_0}} \vert \phi(s, x)\vert^2dx ds \Big] 
$$
$$
\approx \frac{\Delta_C(x_0,y_0)}{\gamma} 
\int_{V_{x_0}\cup V_{y_0}} \vert \psi(x)\vert^2 dx \approx 2 C\Vert \psi\Vert^2
$$
or, for normalized ``wave function'' $\Psi(x),$
$$
\frac{\Delta_C(x_0,y_0)}{\gamma} [ \int_{V_{x_0}} \vert \Psi(x)\vert^2 dx + 
\int_{V_{y_0}} \vert \Psi(x)\vert^2 dx]
$$
$$ =   \frac{\Delta_C(x_0,y_0)}{\gamma} [P(x\in V_{x_0}) +
P(x\in V_{y_0}] \approx 2 C.
$$ 
Hence, during the period of time $T$ there will be produced the following number of double clicks
$$
n_{\rm{double\; click}} = \frac{T\gamma}{\Delta_C(x_0,y_0)} 
\approx \frac{T}{2C} [P(x \in V_{x_0}) +
P(x \in V_{y_0})]\leq \frac{T}{2C}.
$$
Hence, by increasing the calibration constant $C$ one is able to decrease the number of double clicks
to negligibly small. 

\section{Nonobjectivity and contextuality of classical signal theory and quantum mechanics}

Although the probability of double clicks can be made very small, they are fundamentally irreducible. This is one of the reasons why it is impossible to use the functional (as opposed to operator) representation of quantum observables. However, the main reason for this situation is Bohr's contextuality. A classical signal has no sharp position in space, i.e., the (VD) postulate is not valid for classical signals. ``Signal's position'' $x_0$ has meaning only in the context of the position measurement. I note that the scheme of the position measurement described in this paper can be easily generalized to other quantum observables, see [39]. 
In fact, 
$\phi(x_0)$ can be written as $\langle \phi, e_{x_0} \rangle,$ where $e_{x_0}(x)= \delta(x-x_0).$
We can proceed in the same way by taking any basis $e_j$  in the space of signals, instead of the basis consisting of d-functions and corresponding the position measurement.

	Finally, I note that, in addition to Bohr's ontology contextuality, our detection scheme contains another type of contextuality. As we have seen, the probabilities do not depend on the detection threshold $\epsilon.$ Hence, the position observable of QM, $\hat{x},$  is represented by a family of detection schemes indexed by $\epsilon.$ For the same signal, by selecting different $\epsilon$-detectors we obtain different instances of detection and different values of the position observable. However, probabilities related to different $\epsilon$-contexts for the position measurement are the same. Hence, in operational formalism, such as that of QM (cf. [44], [45]), all these detection schemes can be encoded by one symbol, the operator $\hat{x}.$  The same can be said for any quantum observable.

\medskip

This paper was written under the support of the grant Mathematical Modeling of Complex Systems 
of Linnaeus University.

\end{document}